\documentclass[aps,prl,twocolumn,superscriptaddress,showpacs]{revtex4}

\usepackage{subfigure}
\usepackage{graphicx}
\usepackage{bm}
\usepackage{amsmath}
\usepackage{float}

\newcommand{\BiSe}{Bi$_2$Se$_3$}

\begin{document}

\title{Evolution of the Fermi Surface of a Doped Topological Insulator With Carrier Concentration}

\author{E. Lahoud}
\affiliation{Physics Department, Technion-Israel Institute of Technology, Haifa 32000, Israel}
\author{ E. Maniv}
\affiliation{Raymond and Beverly Sackler School of Physics and Astronomy, Tel-Aviv University, Tel Aviv, 69978, Israel}
\author{M. Petrushevsky}
\affiliation{Raymond and Beverly Sackler School of Physics and Astronomy, Tel-Aviv University, Tel Aviv, 69978, Israel}
\author{ M. Naamneh}
\affiliation{Physics Department, Technion-Israel Institute of Technology, Haifa 32000, Israel}
\author{ A. Ribak}
\affiliation{Physics Department, Technion-Israel Institute of Technology, Haifa 32000, Israel}
\author{S. Wiedmann}
\affiliation{High Field Magnet Laboratory, Institute for Molecules and Materials, Radboud University Nijmegen, Toernooiveld 7,NL-6525 ED Nijmegen, The Netherlands}
\author{L. Petaccia}
\affiliation{Elettra Sincrotrone Trieste, Strada Statale 14 km 163.5, 34149 Trieste, Italy}
\author{ K.B. Chashka}
\affiliation{Physics Department, Technion-Israel Institute of Technology, Haifa 32000, Israel}
\author{Y. Dagan}
\email[]{yodagan@post.tau.ac.il}\affiliation{Raymond and Beverly Sackler School of Physics and Astronomy, Tel-Aviv University, Tel Aviv, 69978, Israel}
\author{A. Kanigel}
\affiliation{Physics Department, Technion-Israel Institute of Technology, Haifa 32000, Israel}

\begin{abstract}
In an ideal bulk topological-insulator (TI) conducting surface states protected by time reversal symmetry enfold an insulating crystal. However, the archetypical  TI, Bi$_{2}$Se$_{3}$, is actually never insulating; it is in fact a relatively good metal. Nevertheless, it is the most studied system among all the TIs, mainly due to its simple band-structure and large spin-orbit gap. Recently it was shown that copper intercalated  Bi$_{2}$Se$_{3}$ becomes superconducting and it was suggested as a realization of a topological superconductor (TSC). Here we use a combination of techniques that are sensitive to the shape of the Fermi surface (FS): the Shubnikov-de Haas (SdH) effect and angle resolved photoemission spectroscopy (ARPES) to study the evolution of the FS shape with carrier concentration, $n$. We find that as $n$ increases, the FS becomes 2D-like. These results are of crucial importance for understanding the superconducting properties of Cu$_x$Bi$_{2}$Se$_{3}$.
\end{abstract}


\maketitle

A topological superconductor is a state of matter in which the bulk is fully gapped, but gapless surface states host Bogoliubov quasiparticles \cite{QiZhangReview, HasanReview}. Point-contact  experiments have shown the existence of Zero Bias Conductance Peaks (ZBCP) in Cu$_x$Bi$_{2}$Se$_{3}$ \cite{AndoPointContact,KanigelPointContact}, these were interpreted as a signature of Andreev surface bound states that were theoretically  predicated to exist in certain classes of TSCs \cite{FuBergCuxBi2Se3,FuZBCP}.While the topological properties of TIs are set by the band structure and should not depend on the chemical potential \cite{Bi2Se3_band_Nature}, the properties of the superconducting samples are sensitive to the chemical potential and to the shape of the FS \cite{FuBergCuxBi2Se3}.
\par
Bi$_{2}$Se$_{3}$ has carriers in the conduction band even when carefully prepared \cite{BiSeARPES}. These carriers are believed to be the result of Se vacancies which are always present in the material \cite{Cava_Se_Vacancies}.  The carrier concentration can be increased further by Cu intercalation \cite{HorSuperconductivity, CuxBi2Se3_Nature}. The band structure of Bi$_{2}$Se$_{3}$ is three dimensional (3D), i.e., there is substantial electronic dispersion in the k$_z$ direction. On the other hand, the material is layered, cleaves easily, and its resistivity is anisotropic with $\rho_{zz}/\rho_{xx}\simeq 10$ \cite{Kohler1974}. Band structure calculations \cite{Bi2Se3_band_Nature} indicate that the $\Gamma Z$ dispersion is weaker than the $\Gamma L$ dispersion. Early ARPES experiments have shown that the dispersion can be even weaker than the LDA predictions \cite{OLSON_OLD_ARPES}. It is therefore plausible that upon adding charge carriers, the FS
 will grow in an anisotropic fashion, where  k$_F$ along the k$_z$ direction becomes considerably larger than k$_F$ in the k$_x$k$_y$-plane. The FS can therefore change from being a closed spherical FS at low carrier densities into an open clylinder-like FS at high carrier densities. An important question is whether Cu$_x$Bi$_{2}$Se$_{3}$ has a closed or open FS at carrier densities in which the system is superconducting, n $\simeq 10^{20}$ cm$^{-3}$.
\par
The SdH effect probes extrema in the cross section of the FS. According to the Onsager relation the frequency of the magnetoresistance oscillation as a function of inverse magnetic field is $F=\frac{\hbar}{2\pi e}A(\epsilon_F)$ \cite{SdH}. With $A(\epsilon_F)$ being the maximal cross-sectional area of the FS in a plane perpendicular to the magnetic field. By rotating the field with respect to crystal (see Figure \ref{fig:2}a for the configuration used in this experiment) one can map the full momentum dependence of the FS.
\par
Another common method for studying the FS is ARPES, which was found to be an ideal tool for studying the topological-insulators. ARPES allows one to observe directly the surface states even in samples where the transport is dominated by the bulk \cite{FisherBulkBiSe}.
On the other hand, the surface of a topological-insulator is a very complicated environment for photo-emission; the ARPES spectrum consists of contributions from the bulk-bands, surface states and possibly from a confined 2D electron gas \cite{Hofmann}. One way to disentangle these contributions is to scan the photon-energy used in the experiment. Different photon-energies provide information about the dispersion at different k$_z$ values  and allows one to distinguish 2D-like bands from the 3D bulk band.
\par
In this work we use the two powerful experimental tools: SdH and ARPES; their combination allows us to determine the evolution of the FS as a function of carrier concentration. Using ARPES we show that the Dirac surface states exist throughout the carrier concentration range under study. However, only parts of the bulk-band FS can be clearly seen using ARPES due to the photon-energy dependence of the matrix-elements. The SdH effect, in principle, allows a direct mapping of the entire FS but its amplitude depends exponentially on the effective mass. Consequently, it is less sensitive to high band mass regions on the FS.  Despite the limitations of both probes, together they bring deep insight into the shape and properties of the FS.

\begin{figure}
  \includegraphics[width=1\hsize]{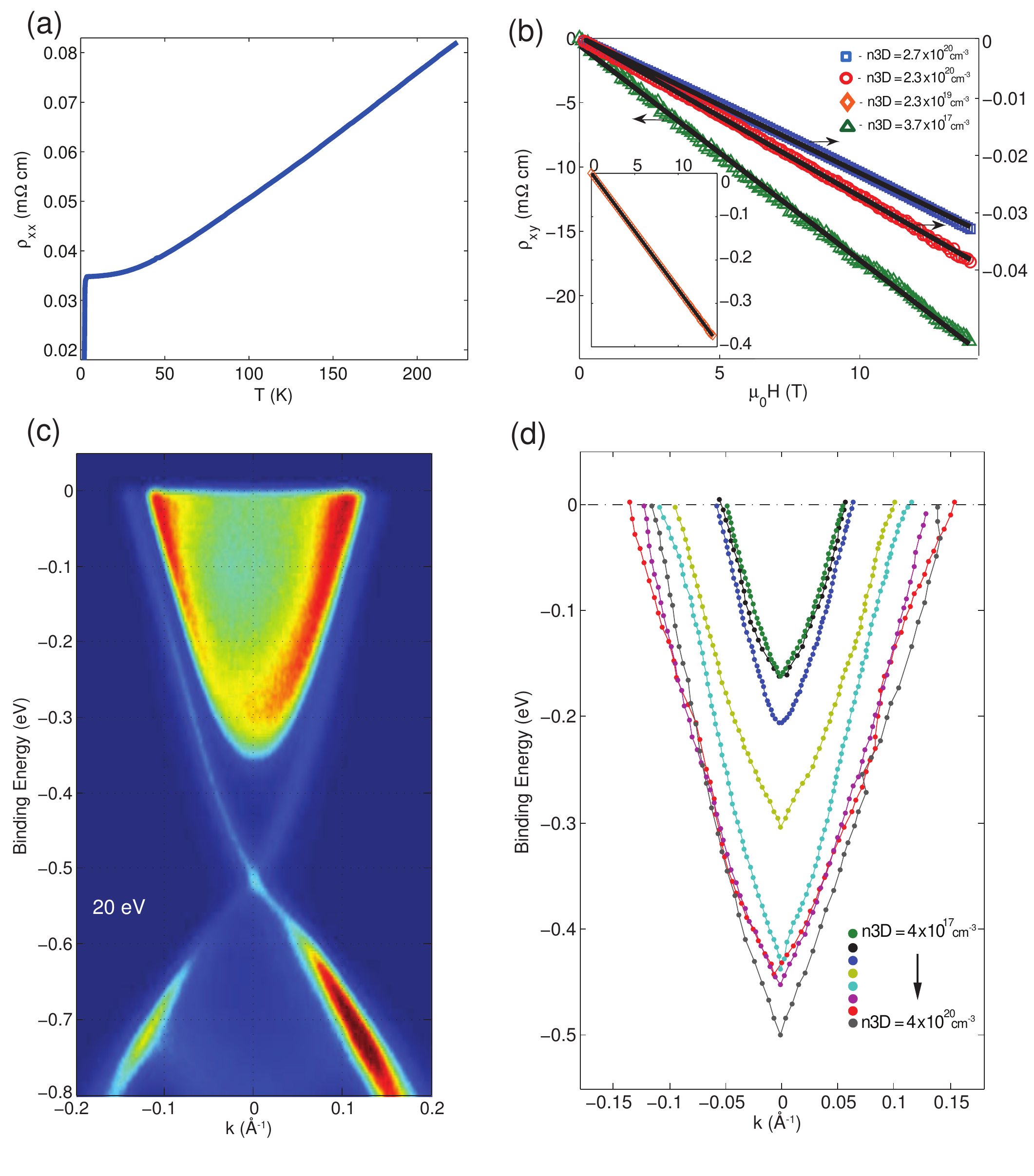}
  \caption{(color online). Transport and ARPES characterization. (a) Longitudinal resistivity versus temperature for Cu doped \BiSe~ with $n\simeq10^{20} cm^{-3}$. Metallic type behavior as well as supercoductivity below $\simeq$ 3K is observed. Different samples may exhibit various superconducting volume fractions and T$_c$ variations. (b) Transverse resistivity versus magnetic field at 2K for three samples. The solid lines are linear fits from which we extract the carrier concentration $n$. Inset: The same for $n\simeq10^{19} cm^{-3}$ . (c) Typical ARPES data from a highly doped sample measured with 20eV photon energy. The detector image shows the dispersion along the $\Gamma$-K direction. One can see that both the surface-state and the bulk-band dispersion coexist. (d) Surface state dispersion as measured using ARPES for 8 different samples with carrier concentrations ranging from 4x10$^{17} cm ^{-3}$ (green curve) to 4x10$^{20}cm^{-3}$ (gray curve).}\label{fig:1}
\end{figure}

\begin{figure*}
\includegraphics[width=1\hsize]{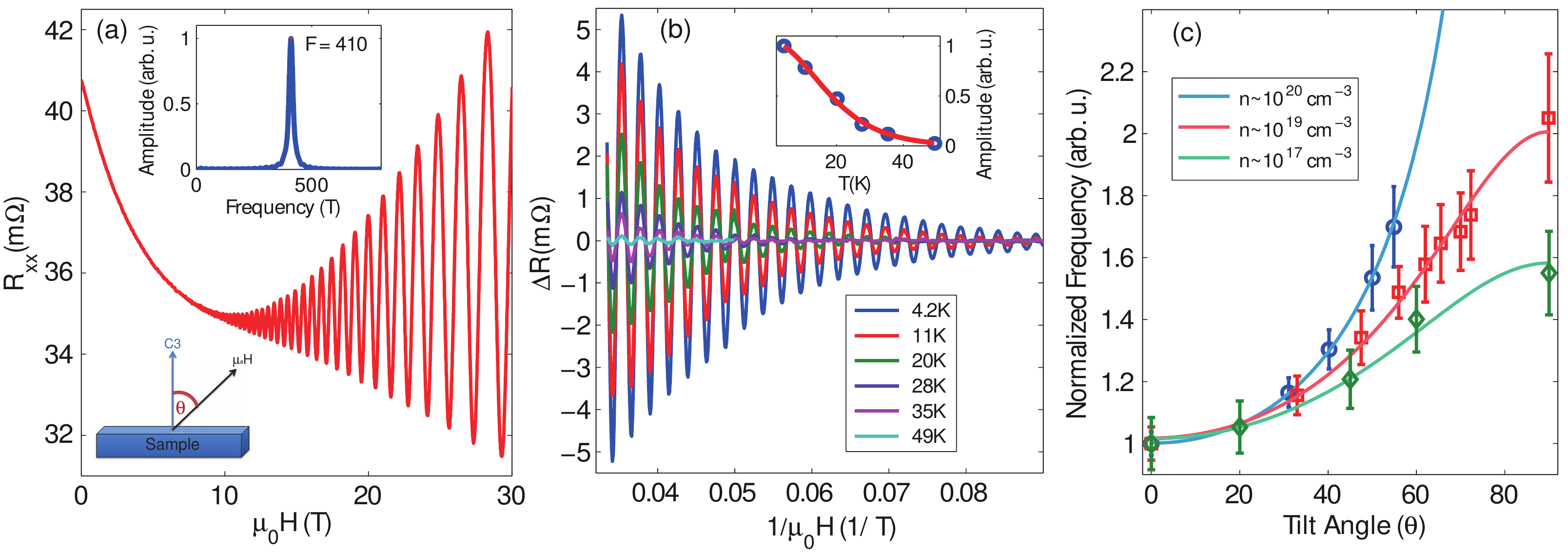}
\caption{(color online). SdH data and analysis. (a) Longitudinal resistance versus magnetic field applied parallel to the C3 axis $(\theta=0)$ at 4.2K for $n\simeq10^{20} cm^{-3}$ (sample A). Inset: Fast Fourier transform (FFT) of these data plotted versus $\frac{1}{\mu_0 H}$ after subtracting a smooth polynomial background. The sharpness of the FFT peak indicates a well defined frequency. Its full width at half maximum is used as an upper limit for the uncertainty in determining the frequency. A drawing of the sample configuration used in this experiment is also shown. (b) Resistance versus $\frac{1}{\mu_0 H}$ after subtraction of a smooth polynomial background at various temperatures (4.2K data are taken from Figure~\ref{fig:2}a). The field is applied parallel to the C3 axis. Inset: Effective Mass is extracted by following the oscillation amplitude at high field as a function of the temperature. The solid line is a fit to the Dingle formula \cite{SdH}, yielding m*$\simeq$$0.24m$$_e$ for this sample ($n\simeq10^{20} cm^{-3}$). (c) The frequency as determined from the FFT versus tilt angle $\theta$ between the magnetic field and the C3 axis for three carrier concentrations. Solid lines are fits for an ellipsoidal FS ($n\simeq10^{17},10^{19} cm^{-3}$) and for a cylindrical FS ($F\propto\frac{1}{cos(\theta)}$) for $n\simeq10^{20} cm^{-3}$ (sample B).}\label{fig:2}
\end{figure*}

\begin{figure*}
   \includegraphics[width=1\hsize]{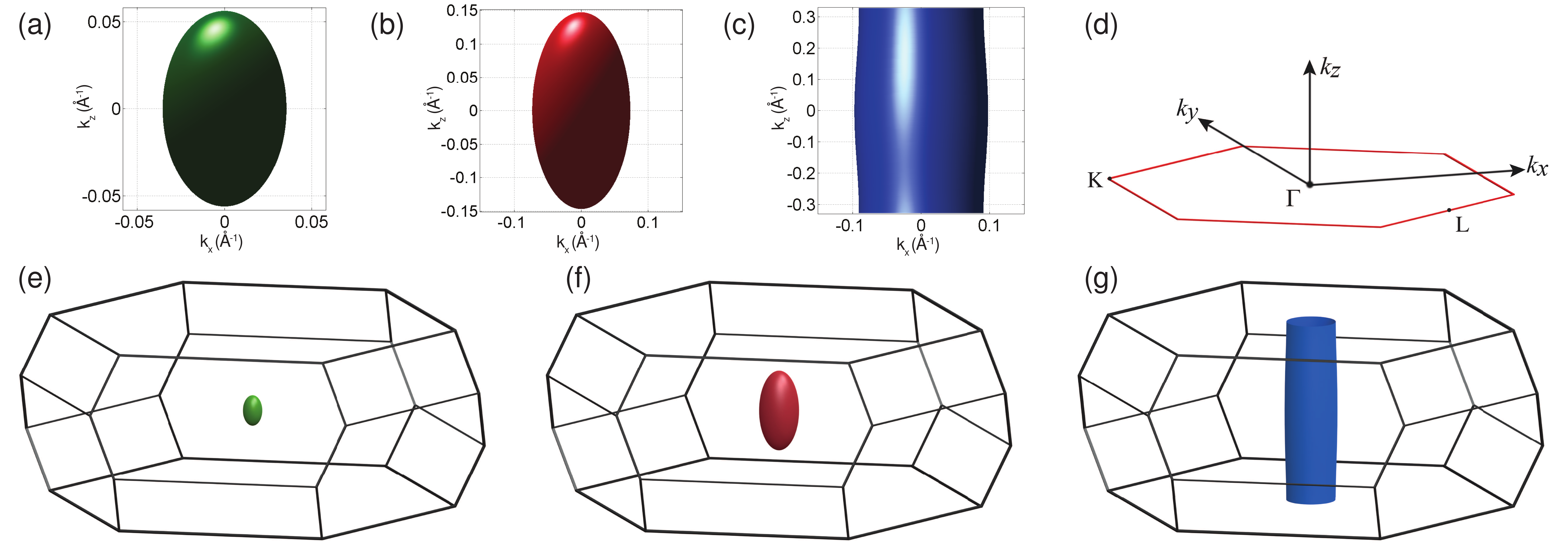}
  \caption{(color online). Evolution of the Fermi surface with carrier concentration. (a) and (b) Calculated ellipsoidal FS from the SdH data in Figure~\ref{fig:2}c, for $n\simeq 10^{17},10^{19} cm^{-3}$ respectively. Detailed profile view of the Fermi surfaces is shown. (c) Calculated FS using tight binding corrugated cylinder model fit to the SdH data in Figure~\ref{fig:2}c (sample B, $n\simeq 10^{20} cm^{-3}$) (see supplementary material for more information). (D) The Brillouin-zone momenta axes. (e)-(g) The Fermi surfaces of (a)-(c) respectively, plotted to scale with respect to the Brillouin-zone.}\label{fig:3}
\end{figure*}

For this experiment, we prepared a series of Bi$_2$Se$_3$ samples with different carrier concentrations. We used two kind of samples: off-stoichiometric Bi$_{2-x}$Se$_{3+y}$ with low carrier density ($n\simeq 10^{17}-10^{19} cm^{-3}$); and Cu intercalated Bi$_2$Se$_3$ with high carrier density ($n\simeq 10^{20} cm^{-3}$). All the samples were prepared using the modified Bridgeman method as described in Ref. \cite{KanigelPointContact,PetrushevskySdH}.
For transport measurements flakes $\sim1-30$ $\mu$m thick were freshly cleaved perpendicular to the C3 axis in a nitrogen environment. Gold contact wires were attached to the samples using silver paint. Hall measurements up to 14 T were performed using DC technique. SdH measurements up to 30 T were performed at the HFML using standard Lock-in technique. The thickness of the flakes was measured in a Scanning Electron Microscope.
The ARPES data was measured at the PGM beam-line at the Synchrotron Radiation Center (SRC), Stoughton WI and at the BaDElPh beam-line at Elettra, Trieste Italy. All the samples were cleaved at base temperature ($\sim$20 K) in a vacuum better than 5$\times 10^{-11}$torr and measured at the same temperature. Each sample was measured for no more than 6 h; within this time we did not observe any change in the chemical potential.
\par
Typical resistivity versus temperature and low temperature Hall measurements are shown in Figures~\ref{fig:1}a and~\ref{fig:1}b, respectively. Typical ARPES data from a Cu intercalated sample with $n \simeq 4 \times 10^{20} cm^{-3}$ is shown in Figure~\ref{fig:1}c. A well defined surface-state can be seen, with a Dirac point at about 500meV below the Fermi-level. The two linearly dispersive surface state branches enclose the parabolic bulk-band whose FS is the subject of this letter. Figure~\ref{fig:1}d shows the dispersion of the surface states for various samples with different carrier concentrations $n$. Upon changing $n$ the Dirac dispersion remains intact with a rigid shift of the Dirac point towards lower energies, while the Fermi velocity (i.e. the slope) remains unchanged.
\par

In Figure~\ref{fig:2} we show SdH data. The resistance as a function of magnetic field for a highly doped sample ($n \simeq 10^{20}cm^{-3}$) is shown in Figure~\ref{fig:2}a. Clear SdH oscillations can be seen. The Fourier transform of these oscillations is shown in the inset. A single, well defined frequency is observed. The oscillations persist up to surprisingly high temperature (see Figure~\ref{fig:2}b). From the temperature dependence of the amplitude the effective mass m$^*\simeq 0.24m_e$ is extracted (see inset of Figure~\ref{fig:2}b).
In Figure~\ref{fig:2}c we show the angular dependence of the SdH frequency for three samples.
For the low carrier concentration samples the oscillations persist up to a tilt angle of $90^\circ$ (see Figure~\ref{fig:2}c), indicative of a closed ellipsoidal FS in. These results are in agreement with previous observations \cite{Kohler1019, Paglione, FisherBulkBiSe, AndoBulkBiSe}. The FS of the $n \simeq 10^{19}cm^{-3}$ is clearly more elongated than the FS of $n \simeq 10^{17}cm^{-3}$.  For the $n\simeq 10^{20} cm^{-3}$ the oscillation amplitude decreases with increasing angle and can not be observed beyond an angle of $55^\circ$. This angular  dependence of the SdH frequency follows almost perfectly  $F\propto\frac{1}{cos(\theta)}$, which is the dependence expected for a cylindrical FS \footnote{This is different than the results in Ref.\cite{PetrushevskySdH}, where the oscillations followed $\frac{1}{cos(\theta)}$ dependence up to $72^\circ$.}.
\par
In Figure~\ref{fig:3} we show the FS of these samples as reconstructed using the SdH data from Figure~\ref{fig:2}c.  The first two samples with $n\simeq 10^{17}, 10^{19} cm^{-3}$, have an ellipsoidal FS. For the third sample ($n\simeq 10^{20} cm^{-3}$) we fit our data to a simplified corrugated-cylinder model (see supplementary material for more information). The resulting FS is shown in Figures~\ref{fig:3}c and~\ref{fig:3}g.
Our SdH data suggest a transition in the shape of the FS from a closed ellipsoid to an open FS as $n$ increases. Below we show that the ARPES data verify this effect.
\par
In an ARPES experiment the signal intensity allows a direct mapping of the electronic dispersion along momentum directions which are parallel to the sample surface.
This is because only the in-plane momentum is conserved. To map the dispersion along k$_Z$, one needs to scan the photon-energy.
 We used the Free-electron final state approximation \cite{Hufner} to find the correspondence between the photon-energy and k$_Z$ (see supplementary material for more information).
\par
\begin{figure}
  \includegraphics[width=1\hsize]{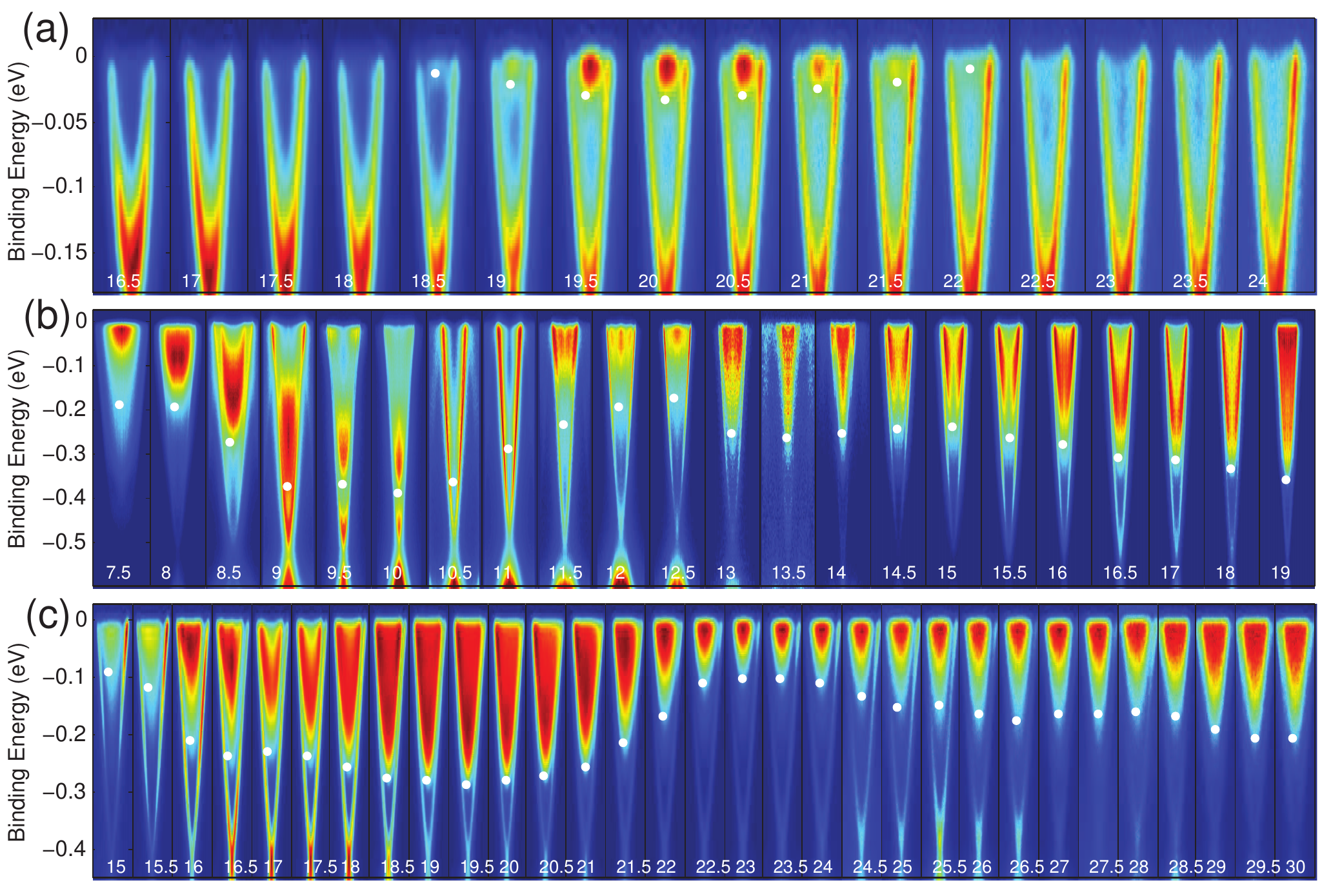}
  \caption{(color online). Photon energy dependence of the ARPES data. We show normal emission data for three different samples: (a) $n=4\times10^{17}cm^{-3}$, (b) $n=4\times10^{20}cm^{-3}$ and (c) $n=2\times10^{20}cm^{-3}$. The white dots represent the bottom of the bulk-band. For the low carrier-density sample the bulk-band is seen only around 19eV ($\Gamma$ point), for the the high carrier-density samples the bulk-band is visible at the entire photon-energy range measured. In panel (b) we show low photon-energy data, believed to be more bulk-sensitive. We find that the bulk-band is visible at the entire photon-energy range measured, which covers a momentum range larger than the $\Gamma$-Z separation.}\label{fig:4}
\end{figure}

In order to map the dispersion along the $k_z$ direction, we performed ARPES measurements at normal emission over a wide range of photon-energies in steps of $0.5$ eV. This was done for various samples with different carrier concentrations. A set of scans is shown in Figure~\ref{fig:4}a for a $n \simeq  10^{17}cm^{-3}$ sample, and in Figure~\ref{fig:4}b and c for two highly doped samples $n \simeq  10^{20}cm^{-3}$.
One can see in Figure~\ref{fig:4} that, as expected, the 2D surface states are insensitive to the photon energy used.
\par

For the low $n$ sample shown, the bulk band is visible only in a narrow range of photon-energies around 20eV which corresponds to the $\Gamma$ point, and completely vanishes as the photon-energy is changed. This indicates that on going along the $\Gamma$-Z direction the dispersion crosses the chemical potential and that the FS is closed. On the other hand, for the high $n$ samples the bulk band remains visible for all photon energies. The maximal width of the bulk-band is obtained at about 20 eV ($\Gamma$ point); the band then disperses upward towards a minimum at both zone boundaries located at photon energies of about $14.5~eV$ and $23.5~eV$. This is a clear indication of an open FS at high $n$.

\par

\begin{figure}
  \includegraphics[width=1\hsize]{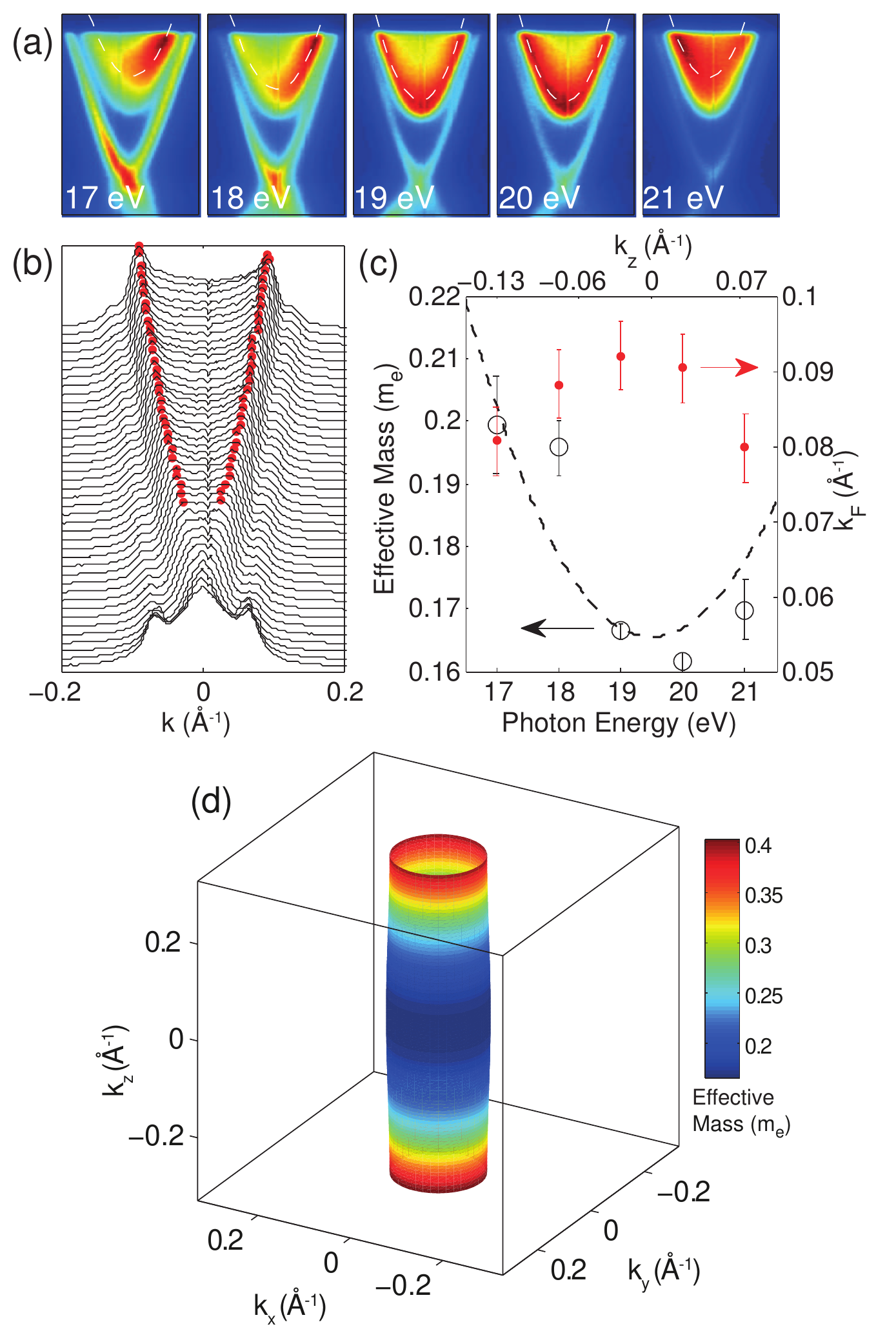}
  \caption{(color online). Effective mass of the bulk-band. (a) Dispersion of the bulk-band around the $\Gamma$-point,  where the dispersion of the bulk is clear and allows an accurate measurement of the effective mass.   The dashed line represents the parabolic fits to the dispersion. (b) MDCs for the 19eV photon-energy data. The red points are the maxima of the MDCs. These maxima are used to extract the dispersion. (c) Summary of the fit results. The red points (black circles) represent k$_F$ (effective-mass) as a function of the photon-energy. (d) A close view of the FS calculated in Figure~\ref{fig:3}c. The color code corresponds to the effective mass calculated for the whole momentum range using the fit in Figure~\ref{fig:5}c. The cylindrical-type shape has a corrugation ratio of $\simeq1.05$. This corrugation ratio together with the enhancement of the effective-mass at the zone boundary explain the absence of a second frequency and the amplitude attenuation at high tilt angles as detailed in the text.}\label{fig:5}
\end{figure}

Next, we look in more detail at the band structure of a $n \simeq  10^{20} cm^{-3}$ sample. Around the $\Gamma$ point the parabolic bulk-band is very clear, this can be seen in Figures~\ref{fig:5}a and~\ref{fig:5}b. By following the peaks in the momentum distribution curves (MDCs), we extract the band dispersion (Figure~\ref{fig:5}b). Fitting the data to a simple parabolic dispersion model we can find k$_F$ and the effective mass at different photon energies. The parabolic best fits are shown as dashed lines in Figure~\ref{fig:5}a. The effective masses resulting from these fits are shown in Figure~\ref{fig:5}c. together with k$_F$. We find that when moving away from the $\Gamma$ point towards the zone-boundary k$_F$ decreases and the effective mass increases substantially.
\par
In Figure~\ref{fig:5}d we show the same FS shown in Figure~\ref{fig:3}c with a color code representing the effective mass , which is measured by ARPES (using the parabolic fit in Figure~\ref{fig:5}c) for various k$_z$. Naively, for this type of FS one would expect two frequencies: the first from the maximal cross-section, at the plane going through the $\Gamma$ point, and the second from the minimal cross-section at zone boundary (the $Z$ points). The k$_z$ dependence of the effective mass extracted from the ARPES data suggests that the SdH signal arising from the minimal-cross section at the zone-boundary will be very weak, as the SdH intensity depends exponentially on the effective mass. This, together with our finding that the area of the cross section perpendicular to k$_z$ changes by merely $5\%$ explain why the second frequency is absent in our measurements. Furthermore, as the angle $\theta$ with respect to the C3 axis increases, the SdH is probing parts of the FS at which the electron's effective mass is larger, so we expect the signal to become weaker, as observed.
\par
The ARPES data suggests a slightly larger corrugation ratio compared to the SdH results, but overall the agreement between the ARPES and the SdH results is impressive. The two probes yield similar values of k$_F$ for the various samples (see supplementary materials), and both techniques show clearly a transition from a closed FS at low carrier concentrations to an open FS at high carrier density. In particular, {\em all} superconducting samples have an open FS.
\par
In order for a time reversal invariant superconductor with odd-parity pairing to be a 3D topological superconductor, it must have a Fermi surface that encloses an odd number of time reversal invariant (TRI) momenta in the Brillouin Zone \cite{FuBergCuxBi2Se3}. We show here that the FS of Cu$_x$Bi$_2$Se$_3$ encloses two TRI points, $\Gamma$ and $Z$.  Our results cast doubt on Cu-doped \BiSe~ as a possible realization of a TRI 3D topological superconductor. Interestingly, this material can be a realization of a 2D-like weak topological SC. Such a system is predicted to have counter-propagating edge-states that can produce Andreev bound states but {\it not} on the (001) surface. If this is the case, the observed ZBCPs in recent point contact experiments \cite{AndoPointContact,KanigelPointContact} can be a result of tunneling into crystalline facets exposing surfaces other than the (001) one.
\par
We are grateful to E. Berg for helpful discussions, S. Lerer for help with the numerical evaluation of the FS and V. K. Guduru for help in the HFML. This work was supported by the Israeli Science Foundation. Work at Tel-Aviv university is supported by the Ministry of Science and Technology under contract 3-8667. The Synchrotron Radiation Center is supported by NSF DMR 0084402. Part of this work has been supported by the EuroMagNET II under the EU contract $n^{\circ} 228043$ and by EU contract $n^{\circ} 312284$.
E.L. and E.M. contributed equally to this work.

\bibliographystyle{apsrev}
\bibliography{myBib}

\onecolumngrid
\newpage

\begin{center}
\textbf{Appendix A: Supplementary Material for "Evolution of the Fermi Surface of a Doped Topological Insulator With Carrier Concentration"}
\end{center}

\subsection{Rigid Shift}
The data presented in Figure 1d of the main text are cuts of the surface state in the $\Gamma K$ direction passing through the $\Gamma$ point. The dispersion was obtained by tracking the peak in the momentum distribution curves (MDCs) in the ARPES spectra. The energy at which we find the Dirac point and the Fermi momentum (of the surface state) for each of these samples is plotted in Figure 6. There is a linear relation between these two quantities, which is expected in the case of a rigid shift of the chemical potential in a linearly dispersing band, where the chemical potential is determined by the carrier density. The scatter in the plot is a result of slight misalignment of the different samples.

\begin{figure}[H]
\centering
\includegraphics[width=0.75\hsize]{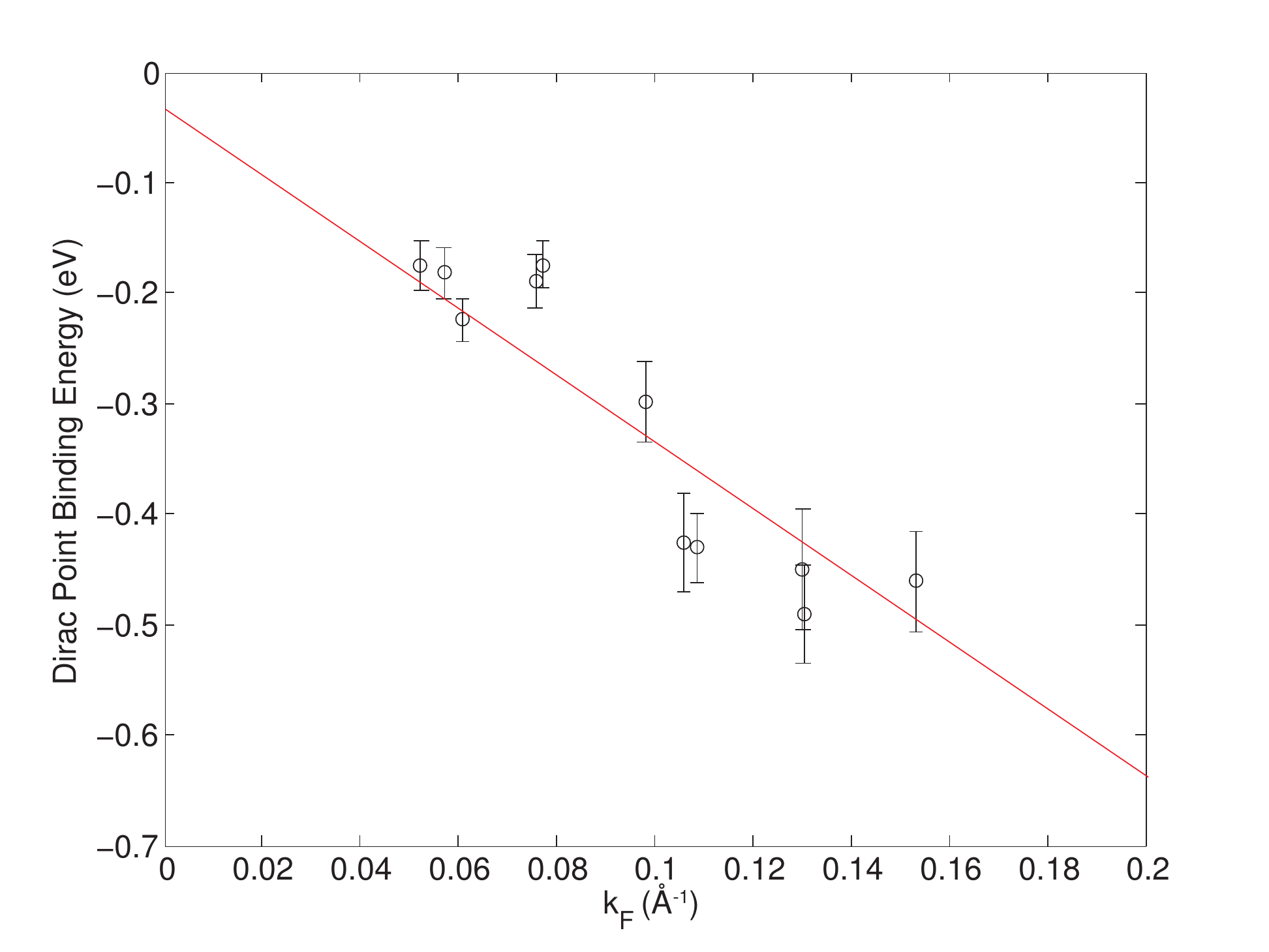}
\caption{The energy position of the Dirac point and the value of $k_F$ for the different samples appearing in Fig. 1d of the main text. Error bars are 95 \% confidence levels.}
\label{fig:supp_Fig1}
\end{figure}

\subsection{Inner Potential}
We relate the value of $k_z$ for electrons inside the crystals to their kinetic energy after photoemission using the free-electron final state approximation $k_z = \sqrt{2m/\hbar^{2} \left( E_{kin} + V_{0} \right)}$. The mapping of band dispersion in the $k_z$ direction is then carried out by changing the photon energy used in the photoemission process, thus changing the value of $E_{kin}$ and obtaining a different value for $k_{z}$. The constant $V_{0}$ is specific to the material and is called the inner potential, formally it is given by $V_{0} = \mu + \phi$, where $\mu$ is the chemical potential measured from the bottom of the band and $\phi$ is the work function. We can determine the value of $V_{0}$ experimentally from the ARPES data, similar to what appears in Fig. 4 of the main text, by identifying the high-symmetry points in the dispersion along $k_{z}$. In Figure 7 we plot the position of the high-symmetry points as a function of photon energy, and obtain a value of approximately $V_{0} \simeq 10.3~eV $.

\begin{figure}[H]
\centering
\includegraphics[width=0.75\hsize]{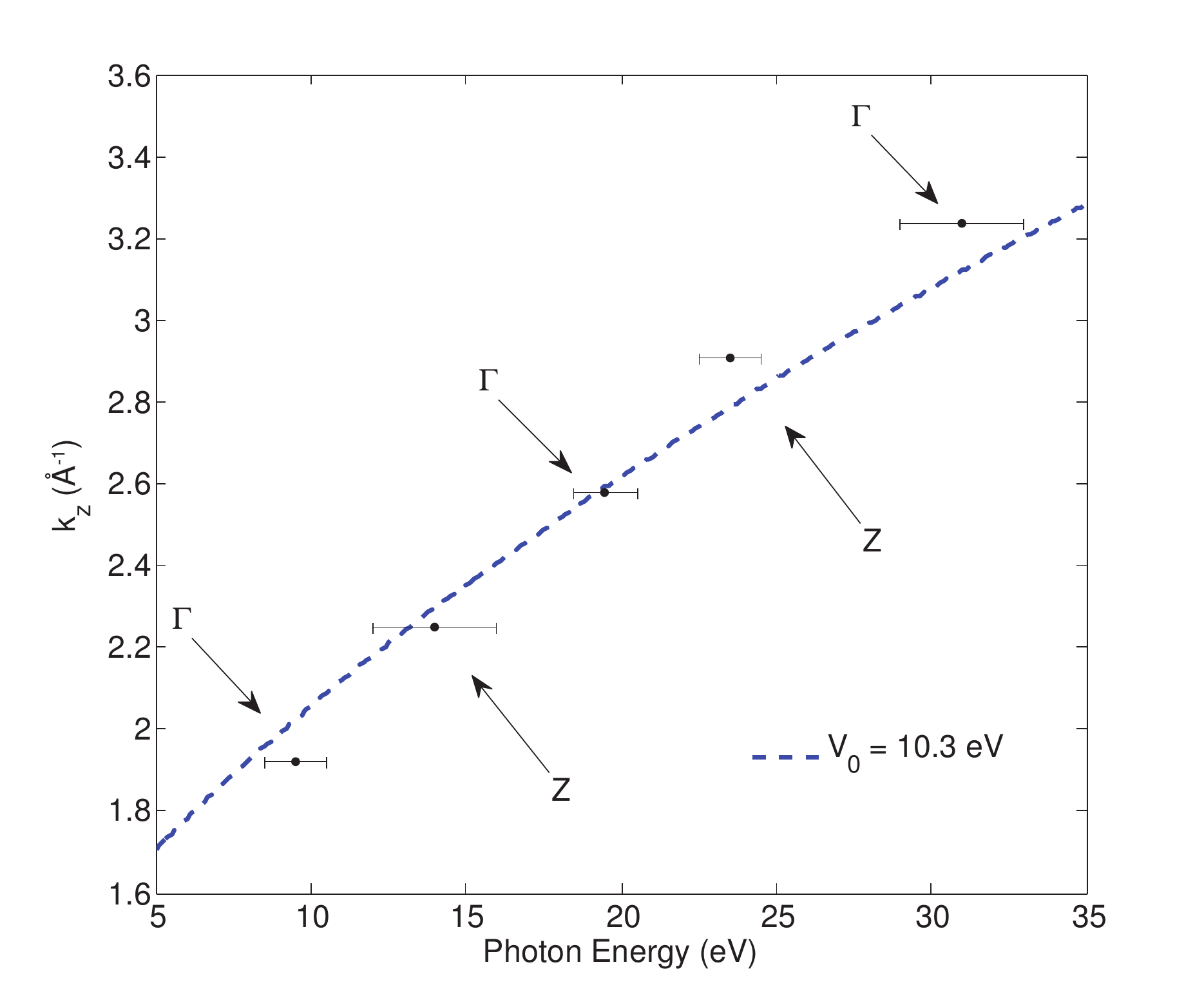}
\caption{Finding the inner potential $V_{0}$: The high symmetry points are identified from the photon energy scans, similar to what is shown in Fig. 4 of the main text. The $k_{z}$ axis separation between the data points is taken to be the $\Gamma$-$Z$ distance, and the error bars is the uncertainty in the position. The dashed line is the equation of free-electron final state approximation (see text) plotted with $V_{0} = 10.3 eV$.}
\label{fig:supp_Fig2}
\end{figure}

\subsection{Band dispersion along $k_{z}$ }
To make a more quantitative analysis we need to find the bottom-of-the-band for each $k_z$ cut. This is found to be a tricky task, for some photon energies a clear parabolic dispersion is seen in the data but for other photon energies we find a parabolic region "filled" with almost constant intensity. We look for the energy for which the intensity drops to half its maximal value; this criterion is used for the entire data set. The $k_z$ values were calculated using the free electron final states assumption with an inner potential $V_0 = 10.3 ~eV$.

In Figure 8, we plot the dispersion of the bulk band along k$_z$, for a few samples with different chemical potential values.
The data for the different samples are plotted relative to the bottom-of-the-band for each sample. The horizontal solid lines represent the chemical potential for each sample measured relative to the bottom of the band.  One can see that the band shifts rigidly with the increasing doping level; this is apparent in the way the data for different samples coincide. The dispersion of the different samples is identical going from the $\Gamma$ point toward the upper Z point. Going from the $\Gamma$ point down, the agreement is not as good. This might be a result of the way we define the bottom-of-the-band. In addition, the free-electron final state approximation can lead to errors.

The dispersion of the sample with the low chemical potential (36 meV) crosses the Fermi-level, an indication of a closed FS. For the other samples we do not find a crossing of the Fermi-level, an indication of an open FS.  Furthermore, for the highly doped samples we find a saturation of the occupied band width. This is again in agreement with an open FS and the absence of a Fermi-crossing point along $\Gamma$-Z.

\begin{figure}[H]
\centering
\includegraphics[width=0.75\hsize]{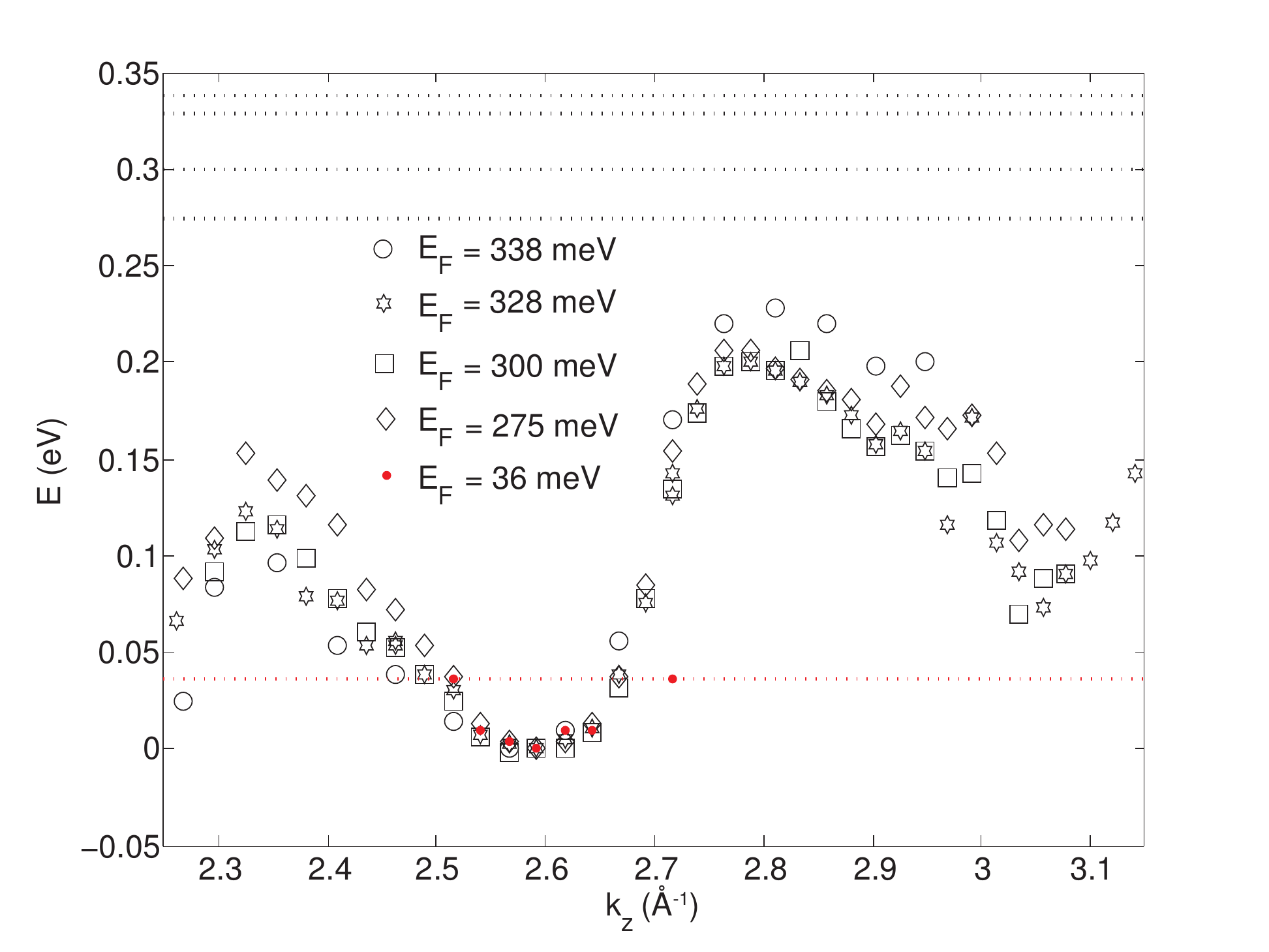}
\caption{Dispersion of the bulk band along the $k_z$ axis. The plot contains data from five different samples with varying Fermi energies (Noted by the horizontal dash lines) as measured from the bottom of the bulk band at the $\Gamma$ point.}
\label{fig:supp_Fig3}
\end{figure}

\subsection{Lower limit for $k_F^z$}
We make a simple estimation of k$_F$ in the $z$ direction, which does not require a measurement of the dispersion along $k_z$.  The surface-state FS always encloses the bulk FS, even for highly doped samples. The k$_F$ we find for the surface-state does not depend on the photon energy. We can use this value as an upper limit for the k$_F$ of the bulk-band in the k$_x$k$_y$-plane. Knowing the carrier density (from Hall measurements), one can calculate the Luttinger FS volume.  Assuming an ellipsoidal FS, $\frac{3\pi^2 n}{k_F^{2}}$ gives a lower limit for k$_F$ in the $k_z$ direction.

In Figure 9, we show the carrier density extracted from the Hall measurements as a function of the surface state k$_F$ for various samples.  The solid line is the carrier density expected for a spherical Fermi surface with a radius corresponding to k$_{F}$. The data points for samples with low carrier density lie slightly below the line, as expected, whereas the points representing high density samples are found above the line. Furthermore, the values we get for k$_F$ in the k$_z$ direction for the samples with $n\sim 10^{20}$ cm$^{-3}$ are larger than the $\Gamma-Z$ distance, indicating that the FS crosses the top zone boundary, meaning that the FS is open.

The blue dots in Figure 9 are the k$_F$ from the SdH data, one can clearly see the good agreement between the values obtained using ARPES and SdH.

\begin{figure}[H]
\centering
\includegraphics[width=0.75\hsize]{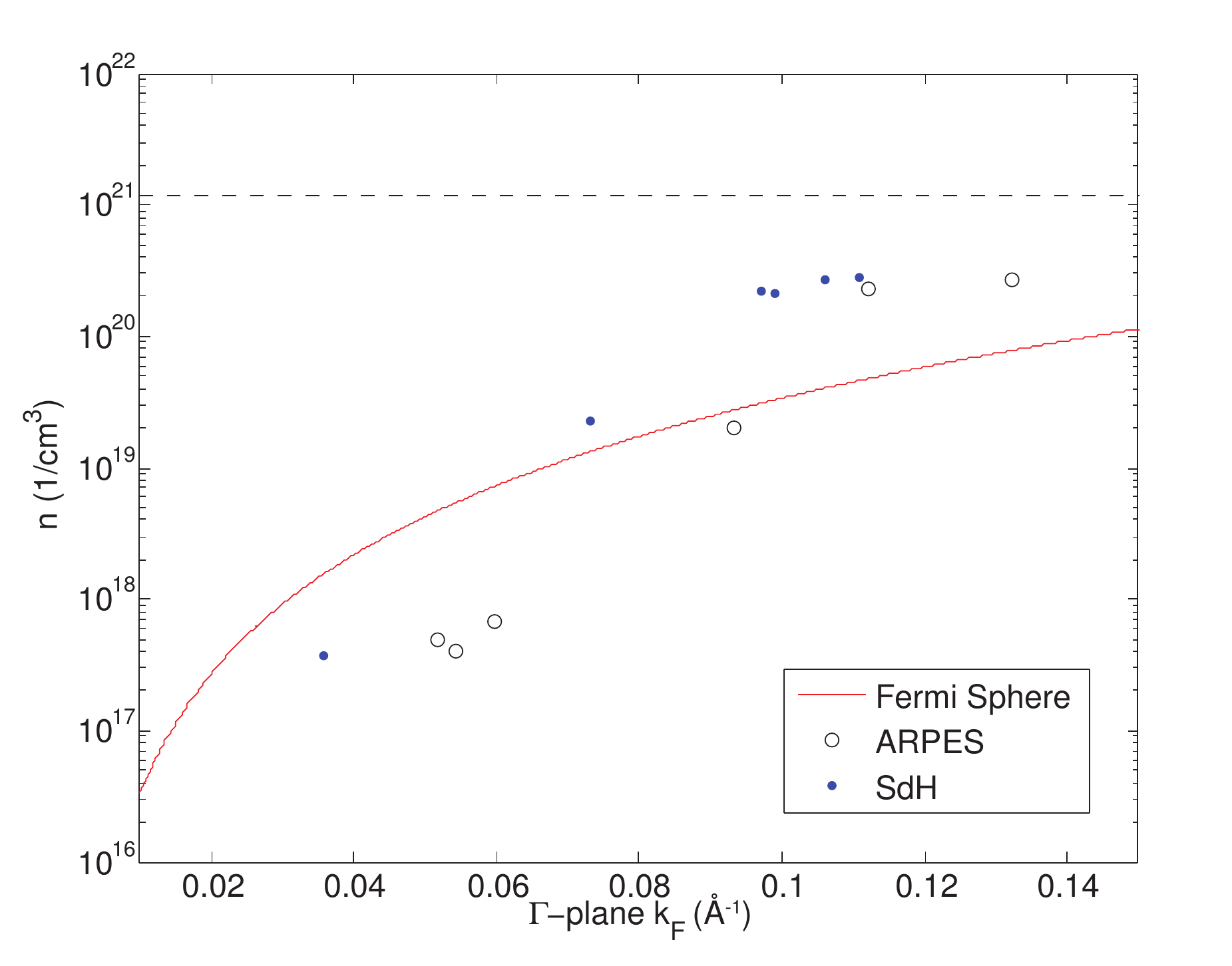}
\caption{Carrier density as a function of $k_F$. The carrier densities are obtained from Hall measurements and they span nearly three orders of magnitude. The $k_F$ (blue dots) was obtained from SdH data, measured with magnetic field parallel to the C3 axis. The $k_F$ (hollow white circles) is that of the surface state obtained from the ARPES data, and it represents an upper bound on the $k_F$ of the bulk. The solid line is the expected carrier density for a spherical Fermi surface. The dashed line is the carrier density for which the radius of the Fermi sphere equals the $\Gamma$-$Z$ distance, that is the Fermi surface reaches the edge of the first Brillouin zone.}
\label{fig:supp_Fig4}
\end{figure}

\bigskip

\subsection{Corrugated Cylinder Model}
We use a simple tight-binding  model to fit the SdH data. Expanding around $k_{x}\simeq k_{y}\simeq0$, we get the equation describing the entire Fermi surface:
\begin{equation}
\frac{1}{2}t_{\perp }\left( k_{x}^{2}+k_{y}^{2}\right) a^{2}+t_{z}\left[
1-\cos \left( k_{z}c\right) \right] =\varepsilon _{F}
\end{equation}
Denotation: a , c - the crystals constants in the $xy$ plane and $z$ axis respectively. $t_\bot$ , $t_z$ - the transfer integrals in the $xy$ plane and $z$ axis respectively, where $t_\bot>t_z>0$.

\par
Applying a magnetic field directing at an angle $\theta$ relative to the $z$ axis and due to symmetry in the $xy$ plane (taking $\phi =0$, where $\phi$ is the angle in the $xy$ plane), the equation for the Fermi surface boundary of the cross sectional region is:
\begin{equation}
\frac{1}{2}t_{\perp }\left( k_{x}^{2}+k_{y}^{2}\right) a^{2}+t_{z}\left[
1-\cos \left( k_{x}c\tan \theta \right) \right] =\varepsilon _{F}
\end{equation}
Measuring wavenumbers in units of $a^{-1}$ and energies in units of $%
t_{\perp }$ this equation becomes:
\begin{eqnarray}
k_{x}^{2}+k_{y}^{2}+2\eta \left[ 1-\cos \left( k_{x}\gamma \tan \theta
\right) \right] &=&2\varepsilon _{F}, \\
\eta &\equiv &\frac{t_{z}}{t_{\perp }}<1,\gamma \equiv \frac{c}{a}>1,  \notag
\end{eqnarray}
The solutions for the cross sectional curve is:
\begin{equation}
k_{y}=\pm \sqrt{2\varepsilon _{F}-k_{x}^{2}-2\eta \left[ 1-\cos \left(
k_{x}\gamma \tan \theta \right) \right] }
\end{equation}
Finally, we calculate the cross sectional area perpendicular to the magnetic field . Using the Onsager relation the corresponding frequency is given by:
\begin{eqnarray}
F\left( \theta,\eta \right) &=&\frac{\hbar }{2\pi
ea^{2}}\int k_{y}d\left( k_{x}/\cos \theta \right) = \\
&&\frac{1}{\cos \theta }\frac{\hbar }{2\pi ea^{2}}\int \sqrt{2\left(
\varepsilon _{F}-\eta \right) -k_{x}^{2}+2\eta \cos \left( k_{x}\gamma \tan
\theta \right) }dk_{x}  \notag
\end{eqnarray}
Expressing the Fermi energy using the zero angle frequency (where the magnetic field is perpendicular to the $xy$ plane):
\begin{eqnarray}
F_{0} &\equiv &F\left( \theta =0,\eta \right)=\frac{\hbar \varepsilon _{F}}{ea^{2}%
}  \notag
\end{eqnarray}
The final equation for the angular dependence of the SdH frequency is:
\begin{eqnarray}
F\left( \theta,\eta \right) &=&\frac{\Delta \left(
\theta,\eta \right) }{\cos \theta }, \\
\Delta \left( \theta,\eta \right) &=&\frac{\hbar }{2\pi ea^{2}}\int
\left\{ 2\left( F_{0}ea^{2}/\hbar -\eta \right) -k_{x}^{2}+2\eta \cos \left(
k_{x}\gamma \tan \theta \right) \right\} ^{1/2}dk_{x}  \notag
\end{eqnarray}

Performing a numerical integration (on $k_x$) combined with a numerical fit (frequency versus magnetic field tilt angle),  one can extract the desired parameter $\eta$ (which defines the corrugation of the cylinder). Our experimental data (the frequency $F$ at different magnetic field tilt angles $\theta$) enables us to perform this analysis. We note that this simple theoretical model does not allow us to find the dependence of the cyclotron mass on $k_z$ as experimentally determined in Figure 5 in the paper.

\subsection{Additional SdH Data and Analysis}
In Figures 10, 11 and 12 we present the SdH measurement of 3 different samples with various carrier concentrations (presented in the paper itself). The measurements are at various tilt angles between the C3 axis and the magnetic field. For these measurements we present: FFT after background substraction for each tilt angle and a fit for each samples Fermi Surface.

\bigskip

\begin{figure}[H]
  \centering
  \includegraphics[width=1\hsize]{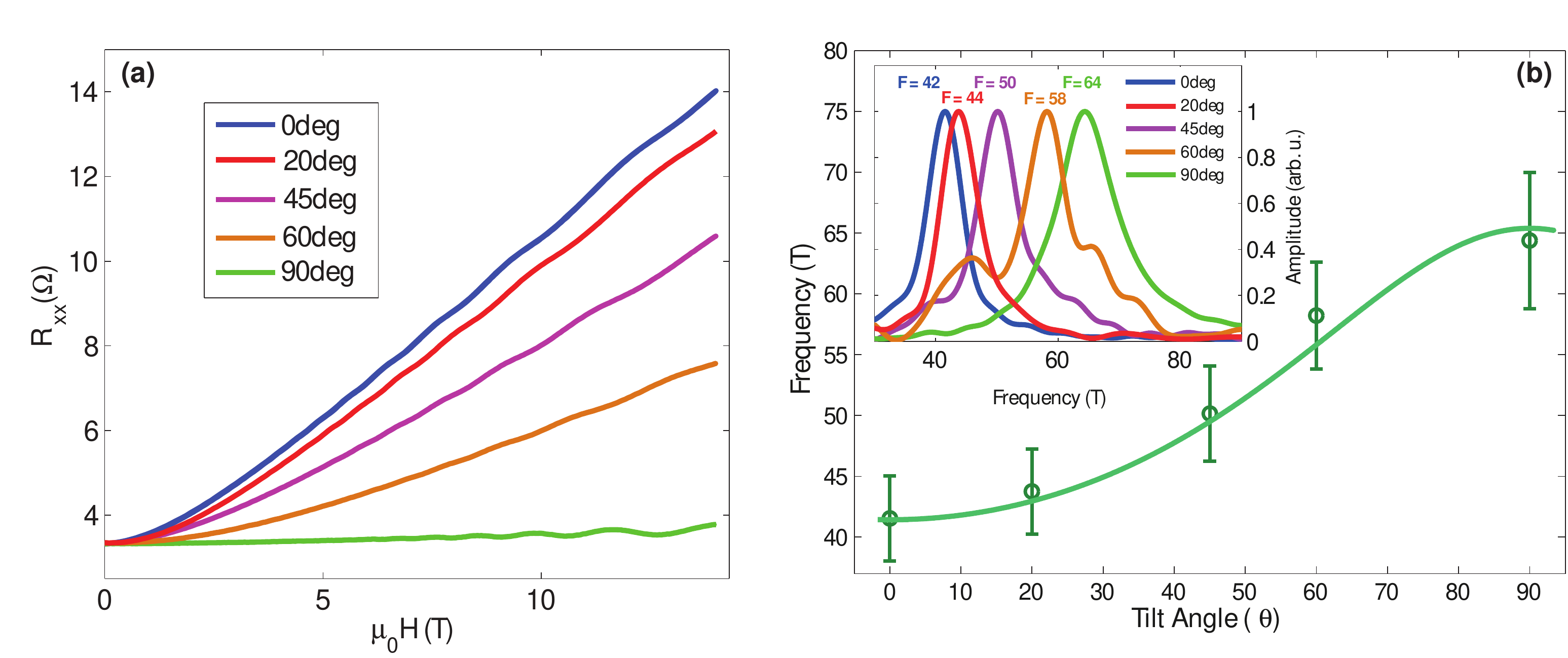}
\caption{(color online). Sample with $n3D\simeq10^{17}$ (a) Longitudinal resistance versus magnetic field at various tilt angles between the C3 axis and the magnetic field . (b) Frequency versus tilt angle. The solid line is a fit for an Ellipsoidal Fermi Surface model. Inset: FFT analysis of each tilt angle.}
\label{fig:supp_Fig5}
\end{figure}

\begin{figure}[H]
  \centering
  \includegraphics[width=1\hsize]{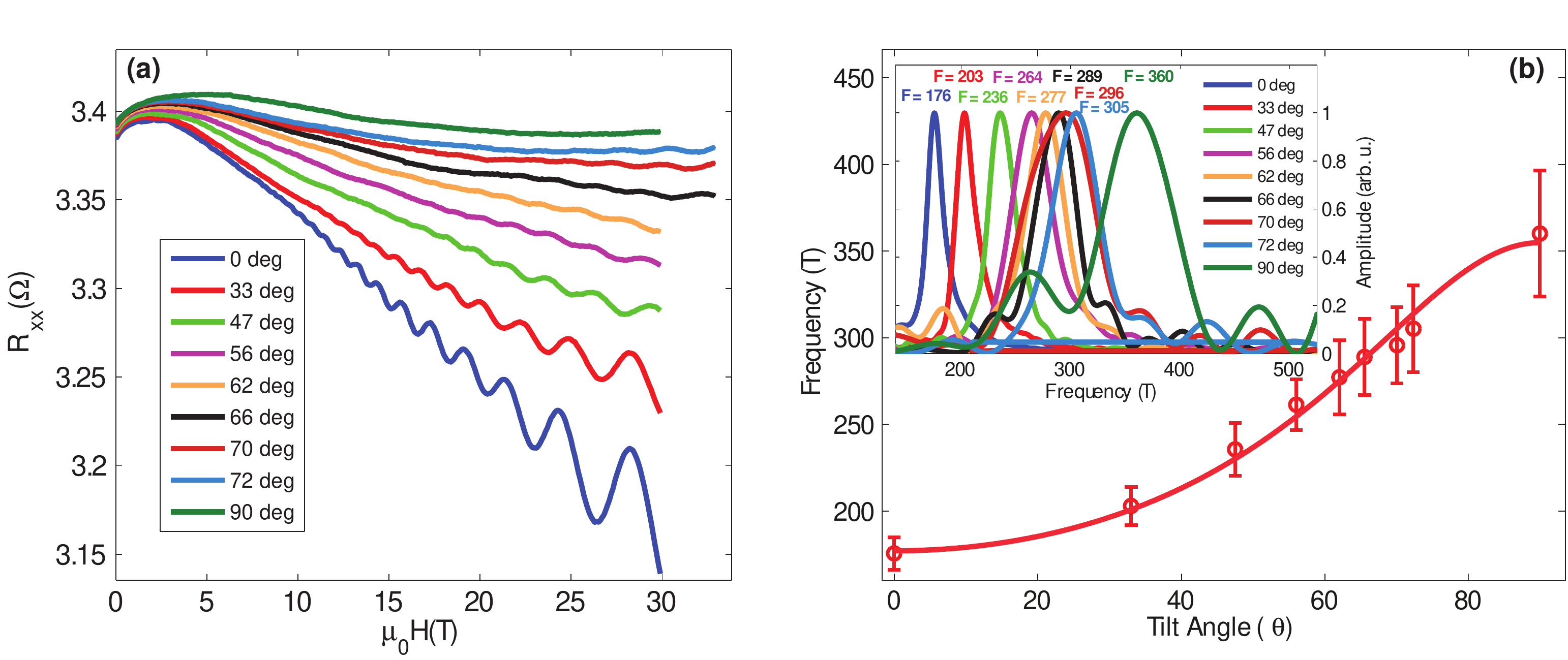}
\caption{(color online). Sample with $n3D\simeq10^{19}$ (a) Longitudinal resistance versus magnetic field at various tilt angles. 0$^\circ$ corresponds to magnetic field parallel to the C3 axis. (b) Frequency versus tilt angle. The solid line is a fit for an Ellipsoidal Fermi Surface model. Inset: FFT analysis of each tilt angle.}\label{fig:supp_Fig6}
\end{figure}

\begin{figure}[H]
  \centering
  \includegraphics[width=1\hsize]{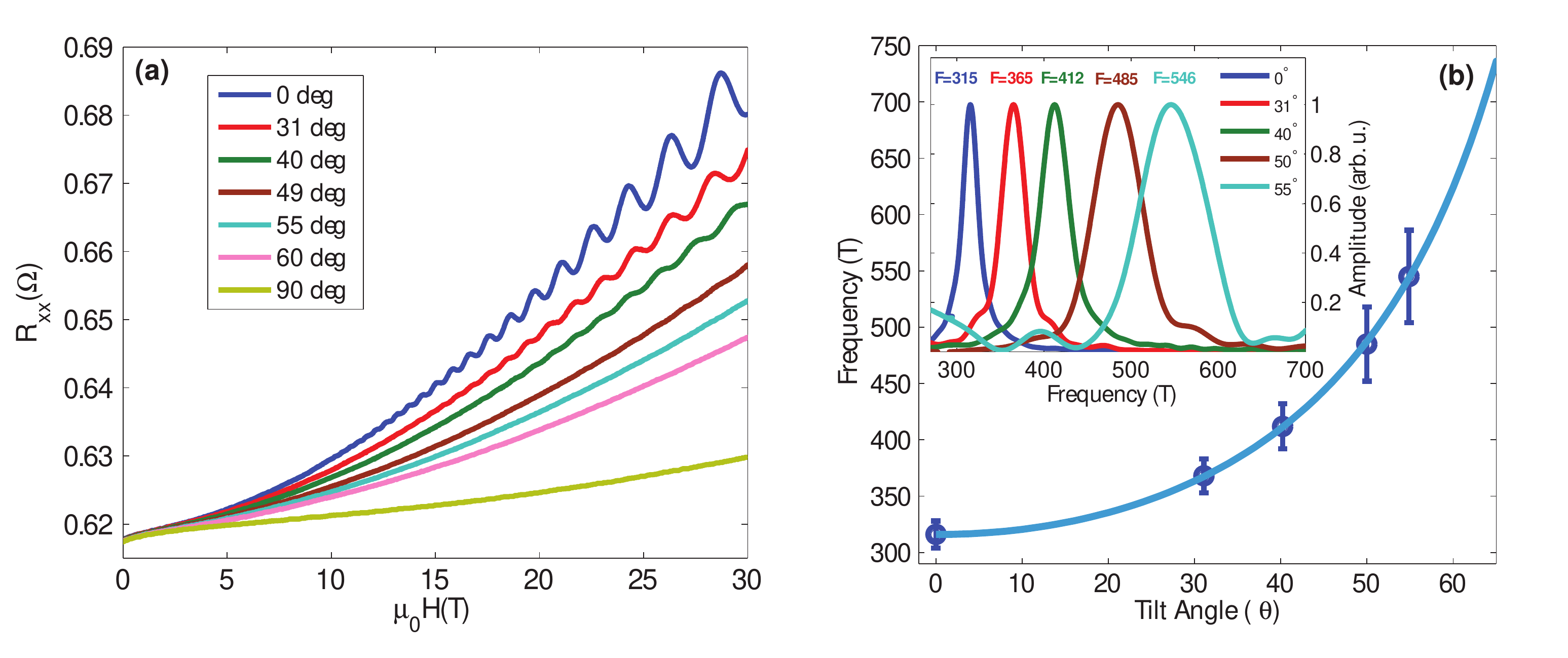}
\caption{(color online). Sample with $n3D\simeq10^{20}$ (a) Longitudinal resistance versus magnetic field at various  tilt angles. (b) Frequency versus tilt angle. The solid line is a fit for a cylindrical Fermi Surface model ($F\propto\frac{1}{cos(\theta)}$). Inset: FFT analysis for each tilt angle.}
  \label{fig:supp_Fig7}
\end{figure}

\end{document}